\documentclass[a4paper,fleqn,usenatbib]{mnras}
\usepackage{mathptmx}
\usepackage[T1]{fontenc}
\usepackage{ae,aecompl}
\usepackage{graphicx}	
\usepackage{amsmath}	
\usepackage{amssymb}	
\usepackage{upgreek}

\usepackage[usenames,dvipsnames]{color} 

\def\ud{{\rm d}}
\def\kms{{\rm km\,s^{-1}}}

\def\msun{{M_{\odot}}}
\title[Stellar multiplicity in high-resolution spectroscopic surveys]{Stellar multiplicity in high-resolution spectroscopic surveys. I. Application to APOGEE subgiants and giants}

\author[Edita Stonkut{\.e} et al.]{Edita~Stonkut{\.e},$^{1,2}$\thanks{{\tt email: edita.stonkute@tfai.vu.lt}} Ross~P.~Church,$^{2}$ Sofia~Feltzing,$^{2}$ and Jennifer~A.~Johnson$^{3}$ \\
$^{1}$Institute of Theoretical Physics and Astronomy, Vilnius University, Saul\.{e}tekio al. 3, LT-10222, Vilnius, Lithuania\\
 $^{2}$Lund Observatory, Department of Astronomy and Theoretical Physics, Box 43, SE-22100, Lund, Sweden\\
$^{3}$Department of Astronomy and Center for Cosmology and AstroParticle
Physics, The Ohio State University, Columbus, OH 43210, USA
}

\date{Accepted 2018.  Received 2018 ; in original form 2018 }
\pubyear{2018}

\begin{document}
\label{firstpage}
\pagerange{\pageref{firstpage}--\pageref{lastpage}}
\maketitle

\begin{abstract}
Many field stars reside in binaries, and the analysis and interpretation of
photometric and spectroscopic surveys must take this into account.  We have
developed a model to predict how binaries influence the scientific results
inferred from large spectroscopic surveys.  Based on the rapid binary evolution
algorithm {\sc bse}, it allows us to model a large, representative population of
binaries and make synthetic observations of observables that would be seen by
the surveys.  We describe this model in detail, and as an application we model
the radial velocity variation of subgiant and giant stars in the Galactic disc,
as observed by the Apache Point Observatory Galactic Evolution Experiment
(APOGEE), part of the Sloan Digital Sky Survey III. Data release 12 of APOGEE
provides an excellent data set for testing our binary models since a large
fraction of the stars in APOGEE have been observed repeatedly.  

We show, by comparing our model to the APOGEE observations, that we can
constrain the initial binary fraction of solar-metallicity stars in the sample to
be $f_{\rm b,0}=0.35\pm0.01$, consistent with the comparable solar neighbourhood
sample.  We find that the binary fraction is higher at lower metallicities,
consistent with other observational studies.  Our model is consistent with the
detailed shape of the high-velocity scatter distribution in APOGEE, which
suggests that most velocity variability above $0.5\,\kms$ comes from binaries.
Our exploration of the binary initial properties shows that APOGEE is mostly
sensitive to binaries with periods between 3 and 3000 years, and is largely
insensitive to the detailed properties of the population.  We can, however, rule
out a population where the mass of the lower-mass star is drawn from the IMF
independently of the more massive star's mass.
\end{abstract}


\begin{figure}
\begin{center}
\includegraphics[width=\columnwidth]{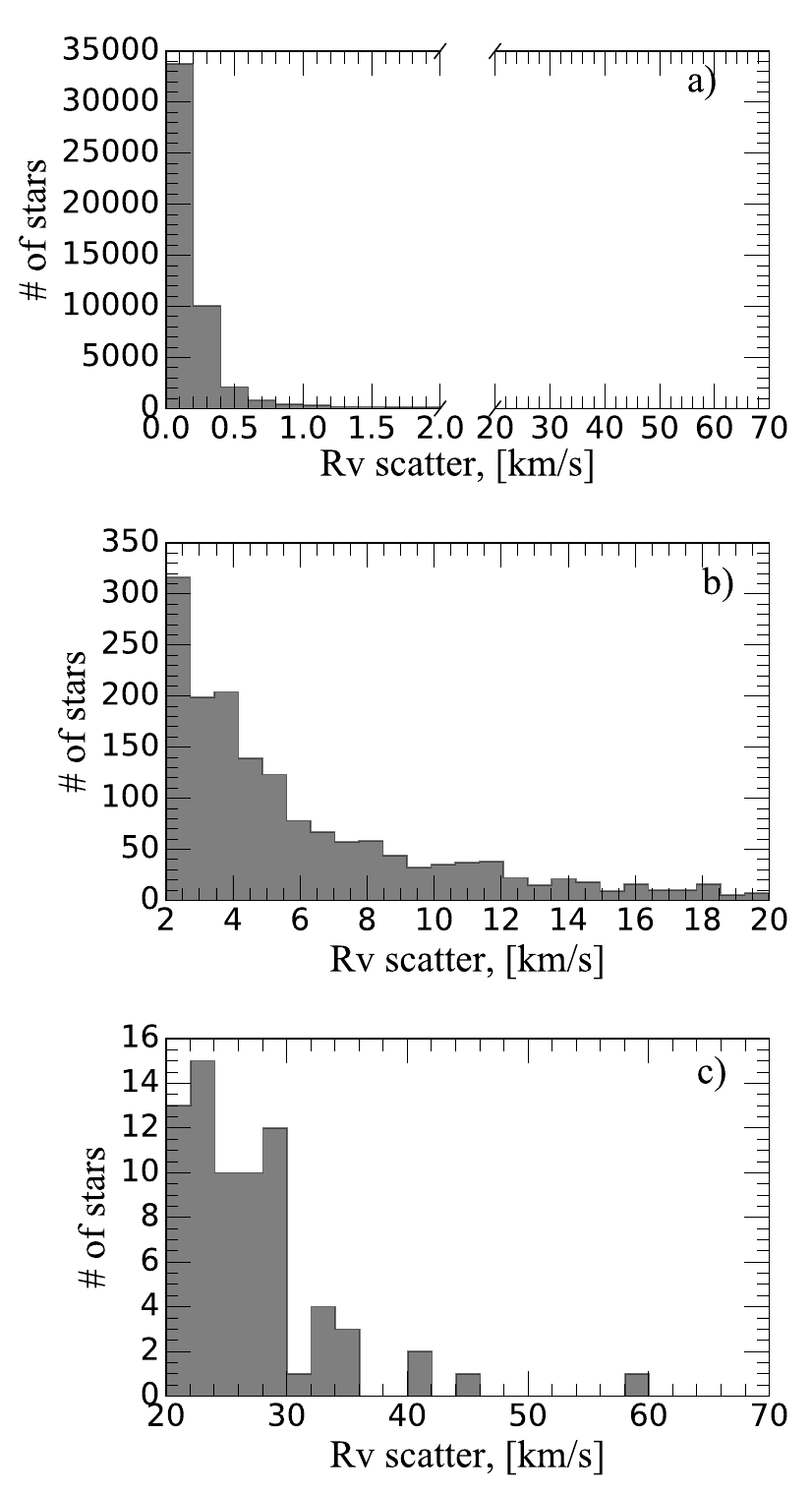}
\end{center}
\caption{
Histograms of the RV scatter of the selected APOGEE Milky Way disc subgiants and
giants in DR12. a) Full distribution of the RV scatter. b) and c) Zooms of
the tail of the RV scatter; note the change in the scale on the $y$-axes.
}
\label{fig:Rvs}
\end{figure}



\section{Introduction}
\label{sec-intro}


Stellar multiplicity is a key parameter for many astrophysical questions.
Several interesting astronomical phenomena, such as gravitational waves
\citep{Abbott16}, gamma-ray bursts \citep{Murguia-Berthier14} and certain types
of supernovae \citep[see][for a review]{Maoz14} arise from binary stars.  For
ongoing and up-coming large spectroscopic surveys, such as RAVE
\citep{Kunder17}, APOGEE \citep{Majewski17}, {\sl{Gaia}}-ESO \citep{Gilmore12},
GALAH \citep{DeSilva15}, LAMOST \citep{Deng12}, 4MOST \citep{deJong16}, MOONS
\citep{Cirasuolo16}, and WEAVE \citep{Dalton16} it is important to identify as
well as quantify the binaries in order to understand the underlying population
and avoid the errors that can arise from mis-identifying binaries as single
stars.  The knowledge of multiplicity also provides constraints on possible
channels of star formation and evolution in the Galaxy.

Surveys of binary stars suggest that the frequency of multiple systems is in the
range of about 22~\% to 80~\%, where the binary fraction is higher for more
massive stars \citep[see][for a review]{Duchene13}.  The most recent study of
binary stars in the solar neighbourhood is by \citet{Raghavan10}.  They updated
and extended the sample of \citet{Duquennoy91}, combining several observational
techniques to search for companions around 454 F6 -- K3 type stars located
within 25 pc of the Sun. They find that 54\% $\pm$ 2\% of solar-type
($\sim$F6--K3) stars in the solar neighbourhood are single, in contrast to the
results of earlier multiplicity studies \citep{Abt76, Duquennoy91}, who found
33\% of solar-type (F3-G2) and 43\% of G-dwarfs are single, respectively.
\citet{fuhrmann11} studied an unbiased, volume-complete sample of more than 300
nearby solar-type stars and showed that than 47\% of stars are single, and that
at least 15\% belong to triple and higher order multiple systems.

Unresolved binaries can appear in spectroscopic observations in two different
forms.  If both stars are of similar brightness in the observed spectral region,
lines from both stars will be visible.  Such a binary is referred to as a
double-lined spectroscopic binary (SB2).  Unless the relative velocities of the
two stars projected onto the line of sight are too similar or the spectral
resolution is low, the Doppler shift owing to the relative motion of the two
stars separates their spectral lines, so these binaries can be recognised from a single
spectrum.  For example, the Geneva-Copenhagen Survey
\citep{Nordstrom04,Holmberg09} identified 19\% spectroscopic binaries in their
$\sim$14\,000 solar neighbourhood F-G type dwarf sample. 

If, however, one of the components in a binary system is significantly fainter
than the other its spectral features are not detected, and we have a
single-lined spectroscopic binary (SB1). SB1s can typically only be identified
via multiple observations.  An example is the RAVE survey, in which 10\% to 15\%
of stars with multiple observations are SB1 candidates \citep{Matijevic11}.
However, not all large-scale surveys have the repeated  observations of stars
that allow us to identify spectroscopic binaries. 

Contamination by binaries has been shown to affect stellar parameters derived
from spectra. In particular, for a low-resolution optical spectroscopic survey
such as SEGUE, the temperature and metallicity of $\sim10\%$ of stars will be
noticeably affected by the presence of an undetected secondary
\citep{Schlesinger10}.  For APOGEE-like spectra of solar-type stars
\citet{el-Badry18} find typical systematic errors of 300\,K in $T_{\rm eff}$ and
0.1 dex in [Fe/H]. They show that binarity leads to larger systematics for
near-infrared than for optical spectra because lower-mass companions contribute
a larger fraction of the total light in the near infrared. 

In this paper we present our model of the effect of binaries on high-resolution
spectroscopic surveys, in order to determine how many binaries will be observed,
whether unresolved binaries will contaminate measurements of chemical
abundances, and how we can use spectroscopic surveys to better constrain the
population of binaries in the Galaxy.  As an application we model binary stars
that mimic subgiants and giants observed by the Apache Point Observatory
Galactic Evolution Experiment (APOGEE) in the Galactic disc. We use the APOGEE
data included as part of the Sloan Digital Sky Survey's 12th data release
\citet{Alam+15}.

\citet{Badenes17} carried out an analysis of binaries in the APOGEE sample,
based on the maximum radial velocity difference $\Delta RV_{\rm max}$ in the
sample of radial velocities determined for an individual star.  They show that
the inferred binary fraction at the present day is a strong function of $\log
g$, consistent with giants in close binaries being depleted by Roche lobe as
they ascend the giant branch.  They also report a strong metallicity dependence
of the binary fraction.  However, they do not model the details of stellar and
binary evolution, which limits the extent to which they can connect the APOGEE
giant population to the birth distributions of the binaries.  Here we carry out
such an analysis.

A brief outline of our paper is as follows. The sample of subgiants and giants
selected from APOGEE is presented in Section~\ref{sec:apogee}. In
Section~\ref{sec:methods} we describe the binary-evolution algorithm and our
choices of initial parameters for the model binary population.  In
section~\ref{sec:SyntheticAPOGEEobservations} we present the synthesis of the
APOGEE observation from models. Then we present our results of binary population
models to examine the effects of binaries on high-resolution spectroscopic
surveys in Section~\ref{sec:Results}. Finally, we summarise our main results,
draw conclusions and explore some perspectives in Section~\ref{sec:Conclusions}.

\section{APOGEE subgiants and giants}
\label{sec:apogee}

APOGEE is a high-resolution ($R \approx 22\,500$), high signal-to-noise
($\approx $ 100/pixel) infrared (1.51--1.70 micron) spectroscopic survey.  The
target selection was defined using the Two Micron All Sky Survey Point Source
Catalog (2MASS PSC) photometry \citep{Skrutskie06}. Stars were selected from the
2MASS PSC employing de-reddened photometry with $H$ band magnitude  $7\leq H
\leq 13.2$.  A simple colour cut of $(J-K_{S})_0 \geq 0.5$ was made to include
stars cool enough for a reliable derivation of stellar parameters and
abundances, and to keep the fraction of nearby dwarf stars in the sample as low
as possible.  Close double or multiple sources are not resolved by 2MASS if
their angular separation is $\leq 6^{\prime\prime}$ whereas the effective
resolution of the 2MASS system is approximately $5^{\prime\prime}$
\citep{Zasowski13}.  APOGEE adopted data quality criteria for targets that the
distance to nearest 2MASS source for $J$, $H$, and $K_{S}$ needs to be $\geq
6^{\prime\prime}$.  The APOGEE fibres are $2^{\prime\prime}$ in diameter, so
this exclusion radius means that no two separate 2MASS sources would be included
on the same fibre.

Data release 12 (DR12 hereafter) was the final product of the third phase of the
Sloan Digital Sky Survey \citep{Eisenstein11} and presented parameters for
146\,000 stars.  The APOGEE Stellar Parameters and Abundance Pipeline
\citep[ASPCAP;][]{GarciaPerez+16} was used to derive $T_{\rm eff}$, $\log g$,
[Fe/H], and elemental abundances by a $\chi^2$ minimisation of the differences
between the observed spectra and a set of synthetic spectra.
\citet{Holtzman+15} compare the temperatures, gravities, and metallicities with
photometric temperatures, seismic gravities, and literature values for open and
globular cluster members, respectively.  They report random uncertainties of
$\approx$ 100\,K for Teff, 0.1 dex for $\log g$, and 0.05 dex for [Fe/H].

The radial velocities (RVs) for each observation are measured as part of the
APOGEE data reduction pipeline \citep{Nidever+15}.  Initial RVs are determined
by cross-correlating the APOGEE spectra with a grid of synthetic spectra. Next,
all the individual observations are shifted and co-added into a combined
spectrum. This combined spectrum is then used as the template to re-derive the
RVs of the individual epochs, again by cross-correlation. \citet{Nidever+15}
found that for the red giants with 3 or more visits and total S/N per pixel
$>20$ the mode of the rms RV scatter was 80\,m/s.  They conclude that this
represents the random uncertainty in APOGEE's RV measurements.
\citet{Badenes17} report a larger typical uncertainty of $0.2\,\kms$.  The
interesting aspect of APOGEE for our work is that the majority of the stars have
been observed multiple times.  APOGEE began observations in May 2011
\citep{Majewski+17} and DR12 contains observations through July 2014
\citep{Alam+15}.  This spread in observation times provides an RV time series
which encodes enough information to detect companions to stars.

\subsection{Star selection}
\label{sec:apogee_selection}

Following the survey targeting strategy for different Galactic components we choose to look at subgiants and giants located in the Galactic disc, where 24$^{\circ}$ $\leq$ $l $ $\leq$ 240$^{\circ}$, $|b|$ $\leq$ 16$^{\circ}$. We select stars for which:\footnote{More documentation on the flags can be found on the CasJobs server at \url{http://skyserver.sdss.org/casjobs/}.}
\begin{enumerate}
\item [(1)] stars.nvists\,>\,2;
\item [(2)] MIN(snr)\,>\,20;
\item [(3)] aspcap.aspcapflag \& dbo.fApogeeAspcapFlag\\(``STAR$\_$BAD") = 0 and aspcap.teff > 0;
\item [(4)] star.starflag \& dbo.fApogeeStarFlag\\(``SUSPECT$\_$RV$\_$COMBINATION") = 0;
\end{enumerate}
This selection finds stars that have been visited more than two times (1) as well where the individual stars' spectra signal-to-noise ratio (S/N) is $\geq$ 20 (2). The next selection step finds a set of stars without any of the flags set that indicate that the observation's RVs (3) or analysis of atmospheric parameters eg., T$\rm{_{eff}}$ (4) is bad.

To avoid contaminating our sample with stars that have their RV
scatter inflated by badly-reduced spectra, 
we select stars using the flag ``SUSPECT\_RV\_COMBINATION$=0$"; i.e. RVs  from
synthetic template will not differ significantly from those from combined
spectrum.  This has the additional effect of removing double-lined spectroscopic
binaries (SB2s) from the sample since these have spectra which are morphologically
different from spectra of single stars.

We select from this potential list of stars only the subgiants and  giants
(where $\log g \leq 4$) with effective temperatures in the range of $3500\,{\rm
K} \le T{_{\rm eff}}\le 5500\,{\rm K}$. In DR\,12 stars have a cutoff
temperature of 3500\,K on the cool side of the spectral grid.  We choose to cut
on stellar parameters rather than photometry since that gives a more reliable
comparison between the models and data.

In order to constrain the Galactic binary population from our model we wish to
model the RV variability of stars that mimic the APOGEE sample and
test whether the observed scatter is comparable with variations induced by
binaries.  We quantify this as the RV scatter, $\sigma_{\rm RV}$,
defined as the standard deviation of RV measurements for stars in the sample
with multiple visits. Hence our final cut is to retain only stars with three or
more RV observations.  This leaves us with a final sample of 49731
stars.  In Figure~\ref{fig:Rvs} we show histograms of the RV scatter of the
selected stars.  The scatter comes from several sources: for example,
instrumental calibration errors, finite telescope resolution, and binary stars.
We discuss this further in Section~\ref{subsec:frequency}.


\section{Binary star evolution models}
\label{sec:methods} 


\subsection{Binary star evolution algorithm ({\sc bse})}

Our modelling of single and binary star evolution is performed using the rapid
binary-star evolution algorithm {\sc bse}, presented in \citet{Hurley02}.  {\sc
bse} enables us to model the evolution of binary systems containing stars of
arbitrary initial mass, metallicity, orbital separation, and eccentricity.  In
addition to all aspects of single-star evolution, {\sc bse} includes mass
transfer and accretion from winds and Roche-lobe overflow, common-envelope
evolution, merger, supernova kicks and angular momentum loss mechanisms.  Orbit
circularization and synchronisation by tidal interactions are included for
convective, radiative and degenerate damping mechanisms.

{\sc bse} is extremely fast, allowing us to model samples of tens of millions of
binaries and hence obtain samples sufficiently large to compare to massive
spectroscopic surveys.  The cost of this is that {\sc bse} is tied to the set of
stellar evolution models that were used to construct it.  We cannot, for
example, incorporate models calculated with alpha abundances enhanced over
scaled solar.  However, this relatively minor limitation is compensated for by
the ease with which we can compute the evolution of large populations of
binaries.

{\sc bse} contains a large number of tunable parameters.  Many of these are not
significant for our calculations -- for example, the treatment of black-hole
natal kicks does not affect our results since we observe vanishingly few
binaries that contain black holes.  For all parameters not described explicitly
below we use the defaults as described in \citet{Hurley02}, a table of which can
be found in Appendix~\ref{app:BSEparams}.

\begin{figure}
\includegraphics[width=\columnwidth]{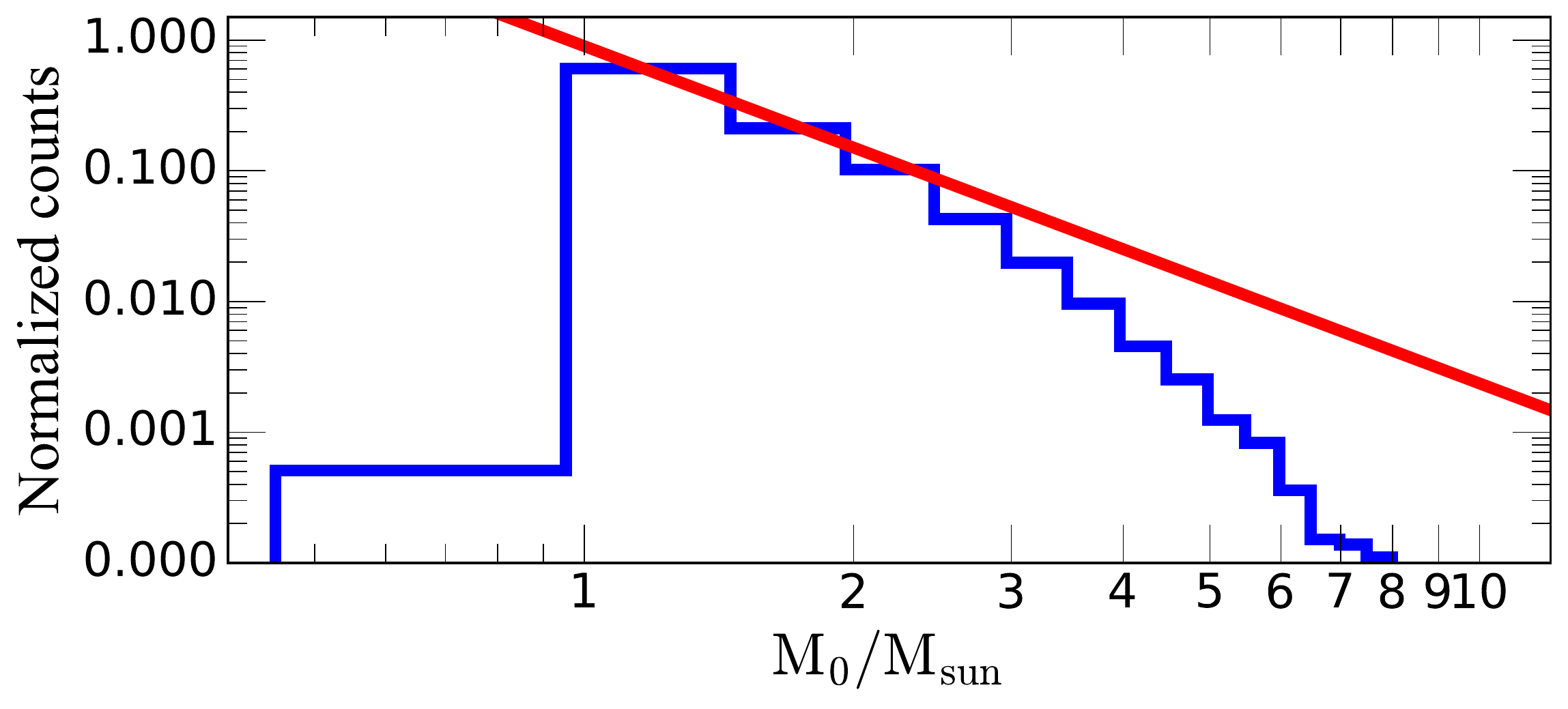}
\caption{
Histogram of the mass distribution for the primary stars in our sample of unresolved
SB1 binaries.   The red line shows the IMF for the initially most massive stars
in our model.
}
\label{fig:param_mass}
\end{figure}

\subsection{Initial parameters for binary population evolution}
\label{subsec:initial} 

In the following sections, we discuss the initial parameters used in our binary
population synthesis, in particular our choice of mass distribution and orbital
parameters.  We distinguish between the initially most massive star, star 1 of
mass $m_1$, and the {\it primary}, which is the most luminous star in the $H$
band at the time of observation.  If star 1 has evolved -- e.g. into a white
dwarf -- by the time of observation, then it may be seen as the {\it secondary}
(less luminous star).

\begin{table}
 \caption{Coefficients of the mass-generating function, Eq.~(1).}
 \begin{center}
 \begin{tabular}{llccccccccc}
  \hline
    \hline
$G_1$ & $G_2$ & $G_3$ & $G_4$\\
  \hline
0.19 & 1.55 & 0.05 & 0.60  \\
  \hline
 \end{tabular}
 \end{center}
\label{tab:G}
\end{table}

\subsubsection{Stellar masses}
\label{subsec:StellarMassesAges} 

We assume that, for each binary, the initially more massive star has a mass
between $0.9\,{\rm M}_{\odot}\le m_1 \le 100\,{\rm M}_{\odot}$. We neglect stars
with primary masses less than $0.9\,{\rm M}_\odot$ as they will not evolve into
giants within the lifetime of the Universe. We generate $m_1$ from the initial
mass function (IMF) of \citet{Kroupa93}, adopting their mass-generating function 
\begin{equation}
m(x)=0.08+\frac{G_1 x^{G_2} + G_3 x^{G_4}}{(1 - x)^{0.58}},
 \label{eq:mass}
\end{equation}
where $x$~follows a continuous uniform distribution on the interval $[0,1]$,
$m(x)$ is in $\msun$, and the coefficients are given in Table \ref{tab:G}.   The
mass of the companion is between $0.1\,{\rm M}_{\odot} \le m_{2} \le 100\,\msun$. The
mass $m_{2}$ of the  companion star is drawn assuming a mass ratio distribution 
\begin{equation}
\rm{f(q) \propto q^\gamma},
\label{eq:gamma}
\end{equation} where the mass ratio $q = m_{2}/m_{1} \le 1$.  In our standard
model $\gamma$ = 0.0; i.e. the distribution is flat.
Figure~\ref{fig:param_mass} shows the IMF for primary stars (red line) and the
histogram of the final mass distribution of observed primary stars in a standard
run.  Above about $1\,\msun$ the distribution roughly follows the IMF, falling
off more rapidly owing to the shorter lifetimes of more massive stars.  Below
about $1\,\msun$ most primaries have not evolved yet, and hence do not appear in
our sample owing to the cut in $\log g$.  There is a small contribution from
stars below the turnoff mass that have accreted from a more massive companion.

\begin{figure*}
\begin{center}
\resizebox{\hsize}{!}{
\includegraphics[width=0.5\columnwidth]{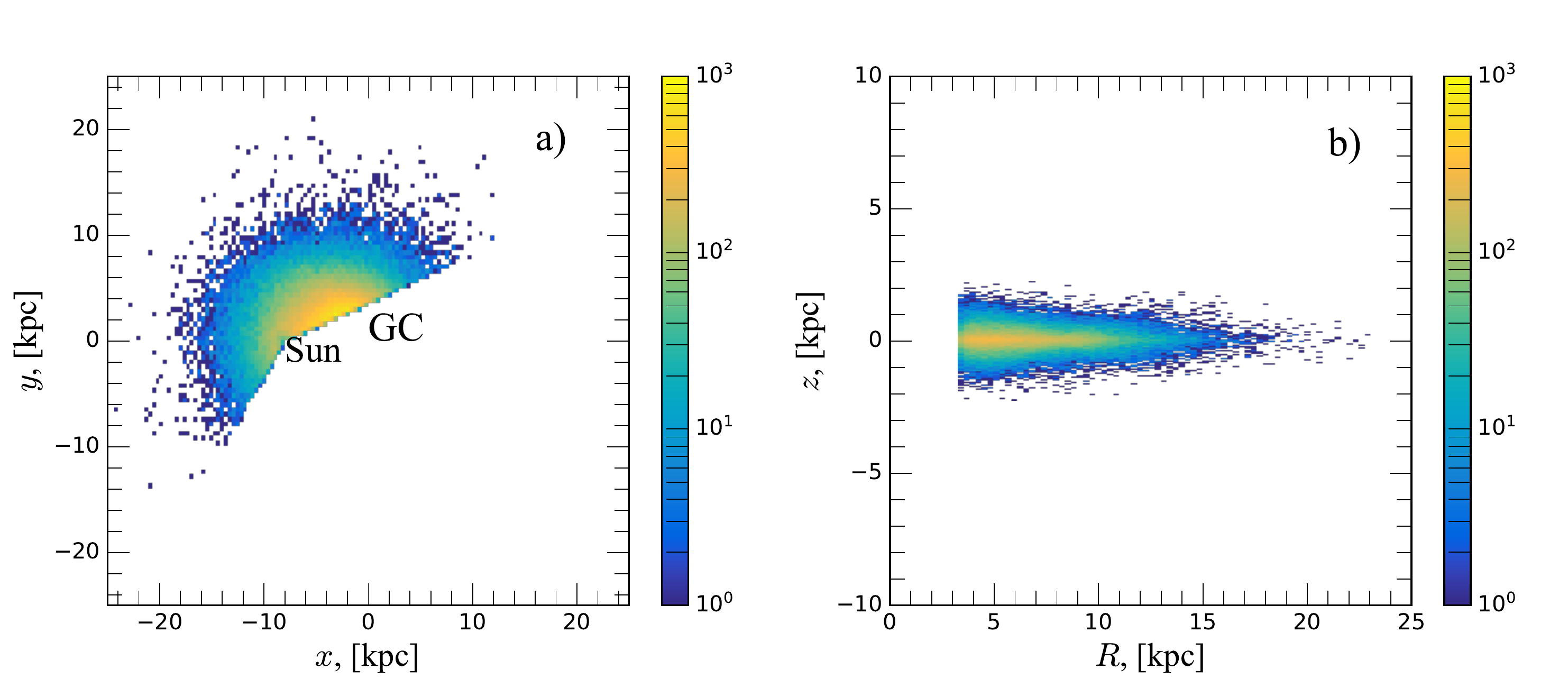}}
\caption{The positions of stars in the Galaxy generated in our model of the stellar disc. On the left: Galactocentric $x-y$ and on the right: $R-z$ plane. To mimic the APOGEE disc sample we only select stars in the model where Galactic longitude 24$^{\circ}$ $\leq$ $l $ $\leq$ 240$^{\circ}$ and Galactic latitude $|b|$ $\leq$ 16$^{\circ}$. The colour coding indicate the number density of stars per bin (as shown in the colour bars).}
\label{fig:model}
\end{center}
\end{figure*} 

\subsubsection{Orbital properties of binary systems}
\label{subsec:binaryProps} 

We assign to each binary pair an initial orbital period in days, $P$, which follows a Gaussian distribution:
\begin{equation}
f(\log P) \propto \exp \Bigg \{ {{\frac{-{(\log P-\overline{\log P})^2}}{2\sigma^2_{\log P}}}} \Bigg \}.
 \label{eq:period}
\end{equation}

\noindent
Here $\overline{\log P}$ = 4.8, equivalent to a peak in the distribution at
$173\,{\rm yrs}$, and $\sigma$$_{\log P}$ = 2.3.  
A similar distribution is found by \citet{Raghavan10} with  $\overline{\log P}$
= 5.03 (293\,years) and $\sigma$$_{\log P}$ = 2.28. 

We allow our binary systems to form with eccentric orbits. The distribution of the orbital eccentricity, $e$, is chosen to be dynamically relaxed; i.e. binaries have all interacted with each other and exchanged energy many times and have reached statistical equilibrium. This thermal distribution thus follows \citep{Heggie75}:

\begin{equation}
f(e) \varpropto 2e.
 \label{eq:ecc}
\end{equation}

Each binary is assigned a random orientation with respect to the plane of the
sky.  In practice this is achieved by taking the longitudes of pericentre and
ascending node to be uniform in $[0,2\uppi]$, and choosing the inclination $i$
so that $\cos i$ is uniform in $[0,1]$.  The mean anomaly is taken to be uniform
in $[0,2\uppi]$: i.e. the fraction of the orbital period that has elapsed since the
last pericentre passage is randomly distributed at the present day.  These four
quantities, together with the stellar masses, orbital period and eccentricity,
determine the initial position and velocity of the two stars.  We convert to a
Cartesian reference frame using a routine taken from {\sc mercury}
\citep{Chambers99}.  The angular separation measured at each observation is then
calculated using the total separation in the $X-Y$ plane.  The RV is taken to be
the value along the $Z$ direction.

\subsubsection{Ages and metallicities}
\label{sec:ageMetallicity}

Since stars in the Galactic disc have a broad age distribution, for our standard
model we assign the age of the binary systems as uniformly distributed from 0 to
10\,Gyr.  As described below, we also test the effect of changing the
age distribution.

{\sc bse} was constructed by fitting functions to models at a fixed set of input
metallicities.  For metallicities between the input values {\sc bse}
interpolates the parameters in the fitting functions.  This leads to significant
errors in the morphology of the giant branch in the interpolated models.  To
avoid this problem we divide the APOGEE observations into bins centred on the
{\sc bse} input model metallicity values.  The binning scheme is summarised in
Table~\ref{tab:binMet}.  Most solar-type stars in the Galactic field are near
solar-metallicity; hence initially we consider the solar metallicity bin. 

\begin{table}
 \caption{Metallicity binning scheme.}
 \begin{center}
 \begin{tabular}{cccc}
    \hline
           &  [Fe/H]     &          & Number    \\
    min    &  model      &   max    & of stars  \\
              \hline                            
    -0.5   &  -0.30      &   -0.15  &    19243  \\
    -0.15  &  0.00       &  0.088   &    18221  \\
     0.088 &  0.17       &  0.45    &     9970  \\
    \hline
 \end{tabular}
 \end{center}
 \label{tab:binMet}
\end{table}

\subsubsection{Description of our models with different initial conditions}\label{sec:Models} 

To model the effects of binaries on high-resolution spectroscopic
surveys, we must have a good understanding of their initial parameters.  The
initial binary parameters are uncertain because of limitations in the different
techniques used to discern them (e.g., difficulties in the observations of the
companion star; incorrect determination of masses because of rotation).
However their properties, e.g.\ primary mass, mass ratio, orbital period,
eccentricity, age, and metallicity, are beginning to be accurately quantified
\citep{Moe17}.  To test the influence of initial conditions on the final result
we construct a variety of models, each with differing but reasonable assumptions
for the initial parameters of the binary population. Table~\ref{tab:models}
lists the models and the parameters we choose to change from the standard setup.
For all the models we synthesise a sample of $N = 2 \times 10^5$ binary stars. 

\begin{table}
 \begin{center}
  \caption{Distributions from which the binary properties are drawn in different
  model sets.}
  \begin{tabular}{llllllllllll}
   \hline
   \noalign{\smallskip}
   Model            &   $m_1$   & $m_2$   & $P$         & $e$ &   ${{\rm Age,~Gyr }}\atop{ {\rm[min; ~max]} }$ \\
   \noalign{\smallskip} \hline  \noalign{\smallskip}
   {\tt standard}   & KTG93$^a$ & $\gamma=0^b$ & DM91$^c$ & t$^e$ &    $0:10$\\
   {\tt m2FromIMF}  &   KTG93   & KTG93        &  DM91    & t     &    $0:10$\\
   {\tt qGamma0.3}  &   KTG93   & $\gamma=0.3$ &  DM91    & t     &    $0:10$\\
   {\tt qGamma1.0}  &   KTG93   & $\gamma=1$   &  DM91    & t     &    $0:10$\\
   {\tt periodR10}  &   KTG93   & $\gamma=0$   &  R10$^d$ & t     &    $0:10$\\
   {\tt flatEcc}    &   KTG93   & $\gamma=0$   &  DM91    & f$^f$ &    $0:10$\\
   {\tt old}        &   KTG93   & $\gamma=0$   &  DM91    & t     &    $10:12$\\
   {\tt young}      &   KTG93   & $\gamma=0$   &  DM91    & t     &    $0:4$\\
   \noalign{\smallskip}
   \hline
  \end{tabular}
 \label{tab:models}
 \end{center}

 {\it Notes:} [Fe/H]=0 for all the models given here; see
 Section~\ref{sec:Zvar}.
 $^a$~KTG93: Mass is drawn from the initial mass function (IMF) from \citet{Kroupa93}. 
    $^b$~Where $\gamma$ is given the mass ratio is selected using
    Equation~\ref{eq:ecc}.  $^c$~DM91: Period distribution from
    \citet{Duquennoy91}.  $^d$~R10: Period distribution from \citet{Raghavan10}.
    $^e$~t:~thermal eccentricity distribution.  $^f$~f:~flat eccentricity
    distribution.
\end{table}

Our baseline model, {\tt standard}, adopts the stellar population properties of
described in \citet{Duquennoy91}.  In {\tt m2FromIMF} we assume that the
secondary star masses, $m_2$, are drawn independently from the same the initial
mass function as the primary, i.e.\ following \citet{Kroupa93}.  Models {\tt
qGamma0.3} and {\tt qGamma1.0} are generated with the mass ratio distribution
following Eq.~\ref{eq:gamma} with ${\gamma}=0.3$ and ${\gamma}=1.0$
respectively.  In model {\tt periodR10} we choose the period distribution
following \citet{Raghavan10}.  The model {\tt flatEcc} has the period
distribution from \citet{Duquennoy91} with a flat eccentricity distribution.  In
the two final models we check the effects of age.  In one model ({\tt old}) the
age of our binary population is exclusively old (from 10 to 12 Gyr).  In the
other one ({\tt young}) we generate a purely young population with binary
systems uniformly distributed from 0 to 4\,Gyr.


\begin{figure*}
\begin{center}
    \includegraphics[width=1.8\columnwidth]{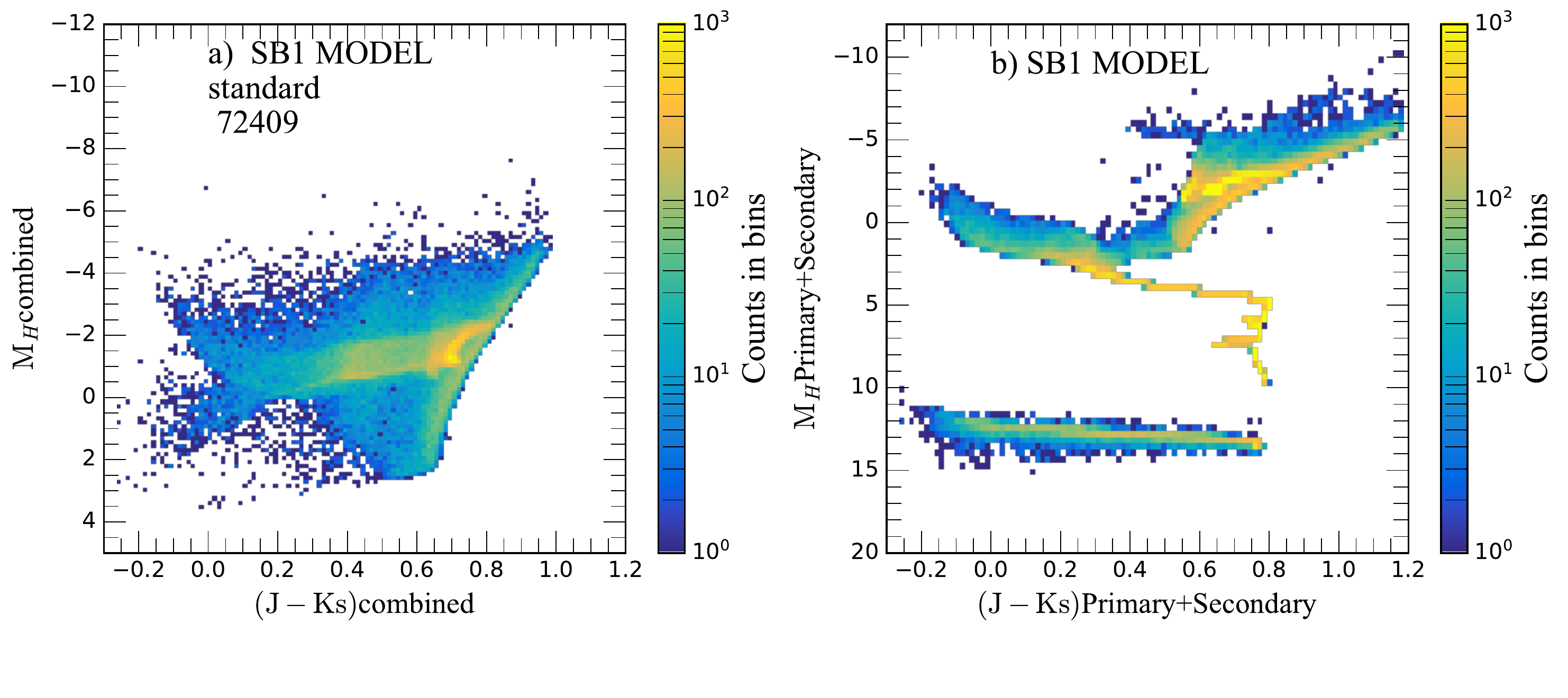}
\caption{
The theoretical colour-magnitude diagrams (CMDs) for single-lined binary stars
(SB1s) in the {\tt standard} model.  {\it a)} The CMD for unresolved binaries,
where the magnitudes and colours show the combined light of both
stars.   {\it b)} the same binaries are shown but magnitudes and colours for
primaries and secondaries are plotted separately; the low mass main-sequence
stars and white dwarfs that appear are secondaries. 
}
\label{fig:CMD}
\end{center}
\end{figure*}


We checked the effects of metallicity on the evolving binary star population by
running  models with [Fe/H]\,=\,[$-$0.30; 0.0; 0.17]; see 
Section \ref{sec:Zvar} for further discussion.   

\subsection{Galaxy model}\label{subsec:galaxy_model} 


To correctly determine the selection of stars we need to know how
far away from the Sun they are.  We distribute the stars according to a model of
the Galactic disc.  We take the disc to be an exponential with radial scale
length $h_{R}=3.0\,{\rm kpc}$ and vertical scale height $h_{z}=0.3\,{\rm kpc}$
\citep{Rix13}. 
In this model the Sun is located 8 kpc from the centre of the Galaxy. We only include the thin disc stars since the APOGEE sample is in the Galactic plane which is dominated by the thin disc.  To mimic the APOGEE disc sample we only select stars with 24$^{\circ}$ $\leq$ $l $ $\leq$ 240$^{\circ}$ and $|b|$ $\leq$ 16$^{\circ}$ (see Fig.~\ref{fig:model}).

\section{Synthetic APOGEE observations}
\label{sec:SyntheticAPOGEEobservations} 

We evolve the binary population using {\sc bse}.  {\sc bse} calculates the
stellar luminosity $L$, radius $R$, mass $M$ etc.~for both stars in the binary
system as they evolve.  In order to mimic the APOGEE observations and to
calculate relative line-of-sight velocities we make multiple synthetic
observations of each model binary.  First, we create an observing schedule for
each binary by randomly selecting one of the actual observing schedules from
APOGEE~DR\,12.  At each observation time we calculate the separation on the sky
(in the $X-Y$ plane) and relative velocity along the line-of-sight (the $Z$
direction).  Other than the effect of the varying orbital phase on the projected
angular separation and RVs we keep the properties of the binary
constant between the observations -- this is reasonable given the long
timescales of stellar evolution compared to the duration of APOGEE observations.  

For a star with a given set of parameters (surface gravity, effective
temperature, mass, radius, metallicity etc.) we calculate  absolute $J$, $H$,
$K_{S}$ magnitudes  and  $(J-K_{S})$ colours in the 2MASS system using
bolometric corrections from \citet{Marigo08}.

To mimic the APOGEE observations we need to  work out how APOGEE would
observe each of our model systems.  A resolved binary star is defined as one
where the two components can be resolved into individual stars. In this study we
define a resolved binary as one where the two stars are separated on the sky by
6$^{\prime\prime}$ or more; see Section~\ref{sec:apogee}.  We analyse such
binaries as two separate stars.  To mimic the selection process for APOGEE we
require that the combined magnitude for a unresolved binary is not too faint
($M_{H}\leq 13.2$) \citep{Zasowski13}.

We define a single-lined spectroscopic binary, SB1, as a binary pair where the
flux ratio of the two components as measured in H-band is $\geq 10$.  A
double-lined spectroscopic binary, SB2, is defined as a binary system where this
flux ratio is  $0.1\leq F_{\rm ratio} \leq 10$. 

Since we are concerned with modelling the red giant stars from APOGEE DR12 we
have chosen to limit our model sample to binary systems where the primary
component has $\log g \leq 4$.  It is straightforward to change this constraint
to suit the observations that are being investigated. 


\section{Results}\label{sec:Results}
 

In this section we first present the results of our model in general terms,
then make a quantitative comparison with the APOGEE data to constrain the
multiplicity frequency.  As we have removed double-lined binaries (SB2s) from
the data (see \ref{sec:apogee_selection}) we consider only the SB1 portion of
the model here.

\subsection{Observed vs. simulated population properties}\label{subsec:effects} 

Synthesised colour-magnitude diagrams (CMDs) of single-lined binaries (SB1) from
the {\tt standard} run are shown in Figure~\ref{fig:CMD}.  The CMD for
unresolved binary stars is shown in panel a). The absolute magnitude is
calculated based on the combined flux of the two stars in the system. The
observed CMD is dominated by red giant branch (RGB) and red clump (RC) stars,
consistent with our selection criteria.
In panel b) we show the same population of binaries but plot the absolute
magnitudes and colours for the primaries and secondaries separately. The
majority of the secondaries are faint white dwarfs or low-mass main-sequence
stars.

\begin{figure}
    \includegraphics[width=1.\columnwidth]{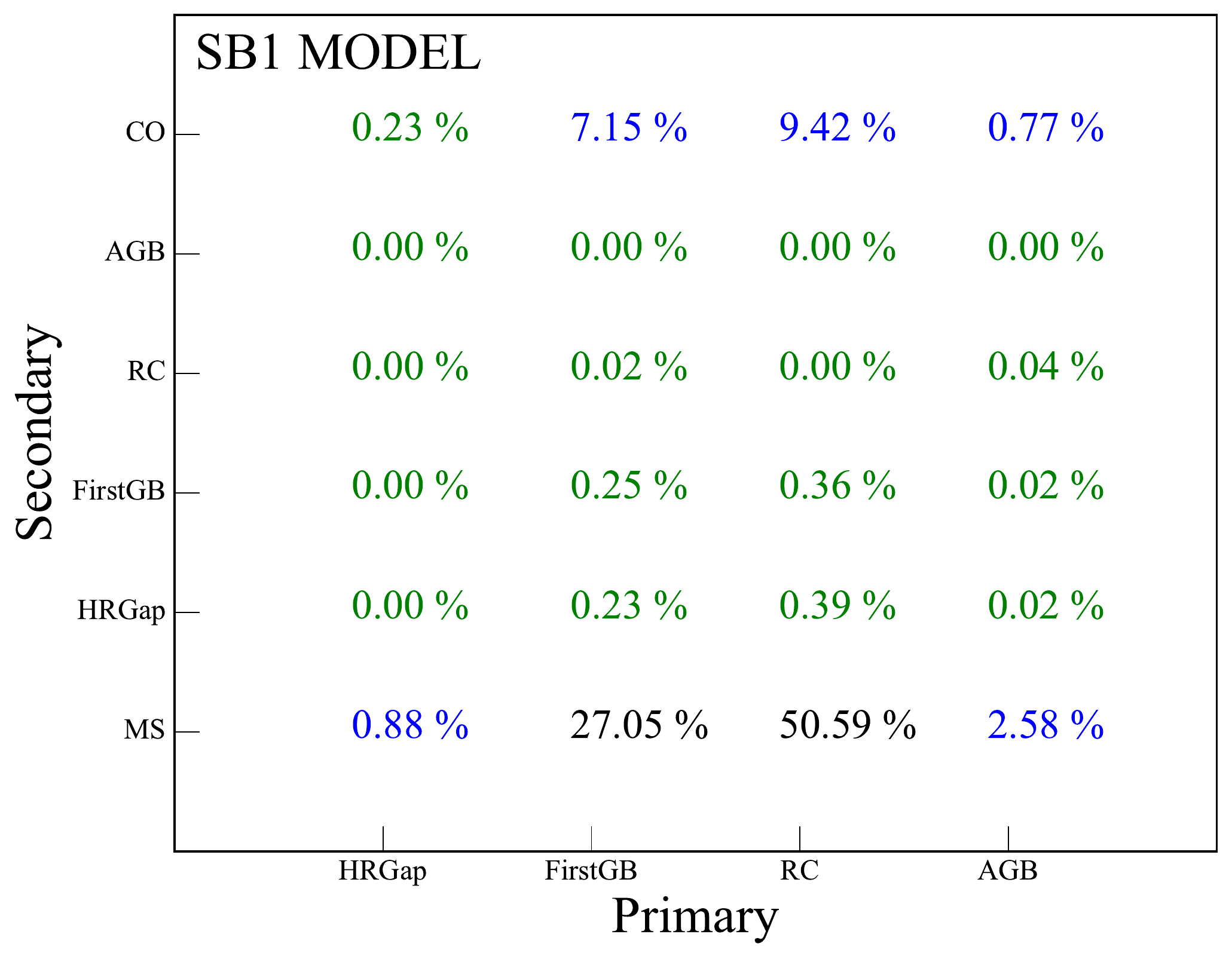}    
\caption{
Stellar types of model binaries observed as single-lined spectroscopic
binaries (SB1) in the {\tt standard} run.  The abscissa shows the type of the
primary (brighter) star and the ordinate the type of the secondary (fainter)
star.  The bulk of binaries have their primary star on the first giant branch or
red clump, with the companions mostly being main-sequence stars.
}
\label{fig:kw}
\end{figure}

The percentage of different stellar types in the {\tt standard} run is
illustrated in Fig.\,\ref{fig:kw}.  The binaries mostly comprise RC primaries
with MS secondaries ($\approx\,50\%$) and RGB primaries with MS secondaries
($\approx\,27\%$). A smaller proportion ($\approx\,17\%$) of these giant
primaries have compact object (CO) secondaries.  There are also small
populations of asymptotic giant branch (AGB) ($\sim\,2\%$) and Hertzsprung
Gap (HRGap) stars ($\sim\,1\%$) as primaries with MS secondaries. 
 
\begin{figure*}
    \includegraphics[width=1.7\columnwidth]{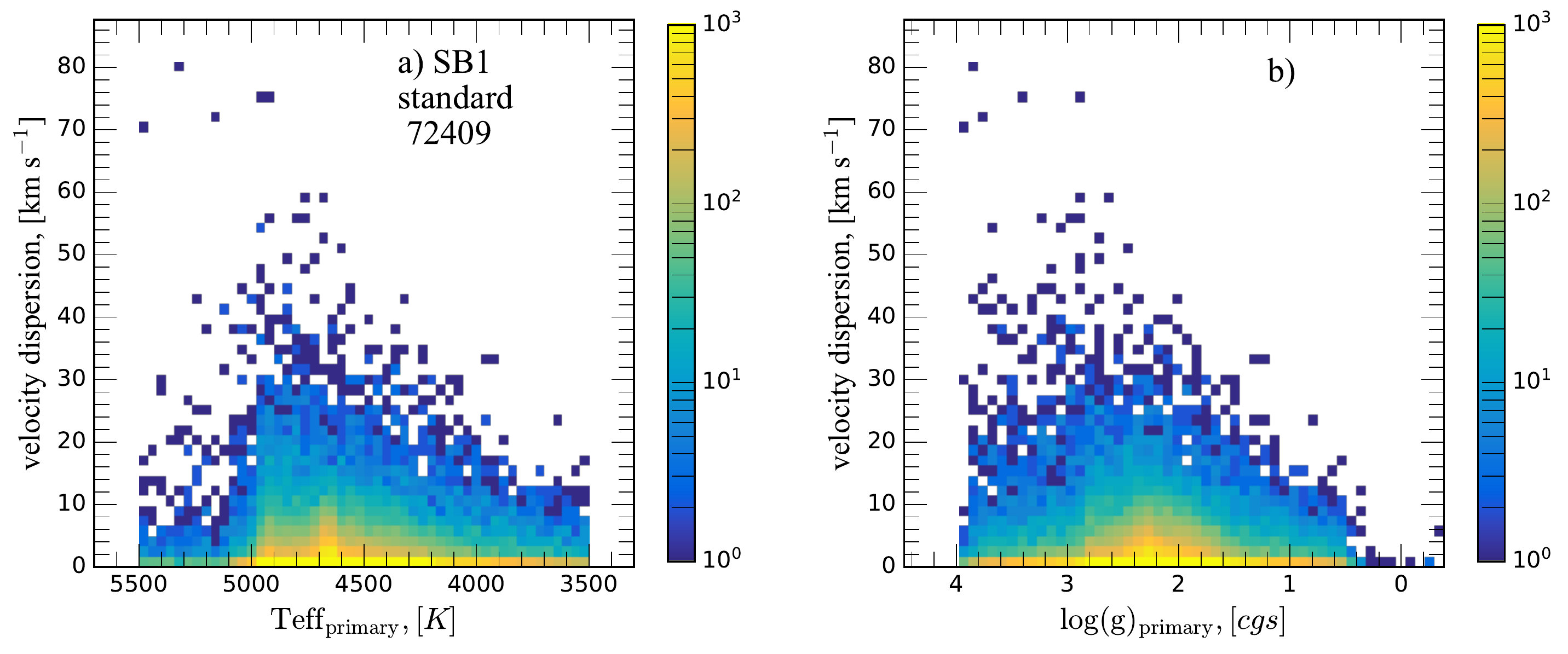}    
    \includegraphics[width=1.7\columnwidth]{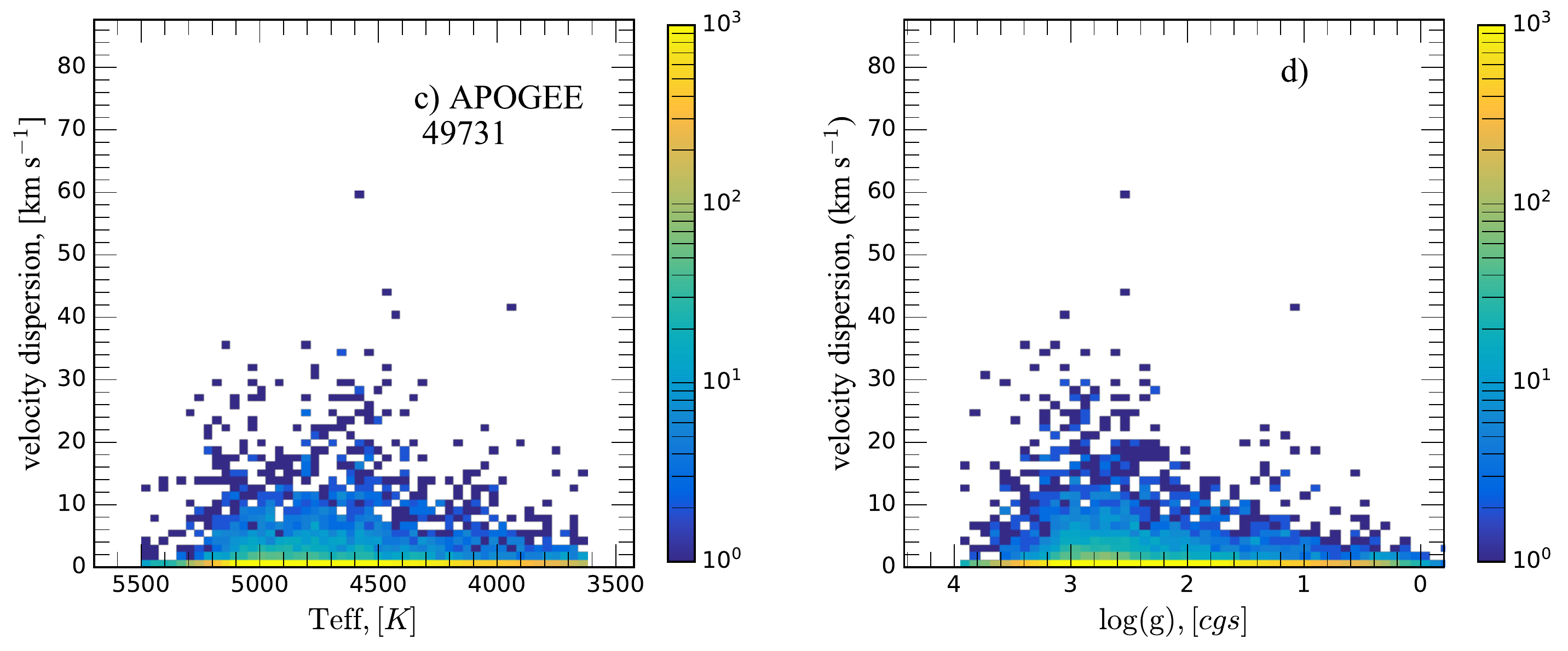}    
\caption{
Velocity scatter versus atmospheric parameters
$T_{\rm eff,Primary}$ and $\log(g)_{\rm Primary}$ from a) and b) the SB1 model, 
and c), d) APOGEE DR~12. The colour coding indicates the number density of
stars.
}
\label{fig:APOGEE}
\end{figure*}

As a first, qualitative comparison between our model and the observed RV
scatter in the APOGEE data we plot the velocity scatter, $\sigma_{\rm RV}$,
versus stellar parameters $T_{\rm eff, Primary}$ and $\log(g)_{\rm Primary}$
(Figure~\ref{fig:APOGEE}). The qualitative agreement between simulated and
observed results is good.  In both model and observations the velocity scatter
decreases as the stars become more evolved: i.e. as $T_{\rm eff}$ and $\log(g)$
decrease.  This is expected, since the star that we observe is the more evolved,
cooler primary.  As stars evolve their radii increase.  Therefore, binary orbits
in which giants can exist must be on average larger than the orbits of
dwarfs: hence the orbital periods are longer and consequently the
binary-induced RV scatter is lower.  The same trend was reported by
\citet{Badenes17} in the analysis of the APOGEE observations.  The plots are for
the {\tt standard} model but similar qualitative behaviour is also seen when the
input distributions are varied.


\subsection{Multiplicity frequency in the field}
\label{subsec:frequency}

Our model allows us to estimate the binary fraction in the Milky Way fields
observed by APOGEE.  We do this by fitting $F\left(\sigma_{\rm RV}\right)$, the
cumulative distribution of the standard deviation of observed RVs
in the APOGEE sample.  We expect that this relationship could vary with
metallicity even though we use the same distributions of birth properties such
as mass ratio.  For example, the range of possible orbital velocities in a
binary is determined by the relationship between stellar masses and radii, which
differs with metallicity.  For the same reasons the relationship in our models
between the initial binary fraction $f_{\rm b,0}$ and the observed binary
fraction at the present day varies with metallicity.  Hence we split the APOGEE
sample up by metallicity.  We choose bins centred on the metallicity values for
which the stellar evolution models that underlie {\sc bse} were constructed,
since the giant branch models that it produces are most reliable at those
metallicity values (see Section~\ref{sec:ageMetallicity}).  Our binning scheme
is given in Table~\ref{tab:binMet}.

The RV variations of the stars in the APOGEE sample come from many
sources.  Some are astrophysical: for example, binaries and higher order stellar
multiple systems, star spots, oscillations of the stellar surfaces, and
planetary systems.  Some are not: for example, errors in instrumental
calibration and data reduction.  We lump together all the RV variations {\it
not} caused by binaries and model them with a single distribution. In practice
we find that this distribution is adequately described by a log-normal
distribution whose mean $\mu_s$ and standard deviation $\sigma_s$ become
nuisance parameters to be fit. The RV variations from binaries we obtain
separately from our model.

In detail, we synthesise a mixed sample of single stars and binaries. At birth,
a fraction $f_{\rm b,0}$ of our objects are binaries, the remaining $1-f_{\rm
b,0}$ being single stars.  In this definition the initial binary fraction is
\begin{equation}
f_{\rm b,0} \equiv \left.\frac{N_{\rm binaries}}{N_{\rm single} + N_{\rm binaries}}\right|_{\rm birth}.
\end{equation}

For each star or binary that we generate we draw from
the appropriate synthesised sample described in
Section~\ref{sec:SyntheticAPOGEEobservations}.  If it does not meet the
selection criteria described in Section~\ref{sec:apogee}, for example because it
is too faint, we discard it.  Otherwise we make a synthetic observation as
follows.  We first obtain an observing schedule by drawing randomly from the
APOGEE visit schedule.  For each visit we draw a velocity from the  log-normal
distribution characterised by $\sigma_s$ and $\mu_s$ to produce a synthetic set
of RVs.  In addition, for binaries, we obtain from our model the
velocity owing to the primary star's motion around the binary's barycentre,
projected onto the line of sight at the visit times, and add them to the values
from the log-normal distribution.  Finally we take the standard deviation to
measure the scatter.  This process is continued until we have made synthetic
observations of the same total number of objects as in our APOGEE sample.  We
then sort by scatter to obtain the cumulative distribution of RV
scatter in the model, which we denote $F_{\rm mod}(\sigma_{\rm RV})$.

\begin{figure}
\includegraphics[width=\columnwidth]{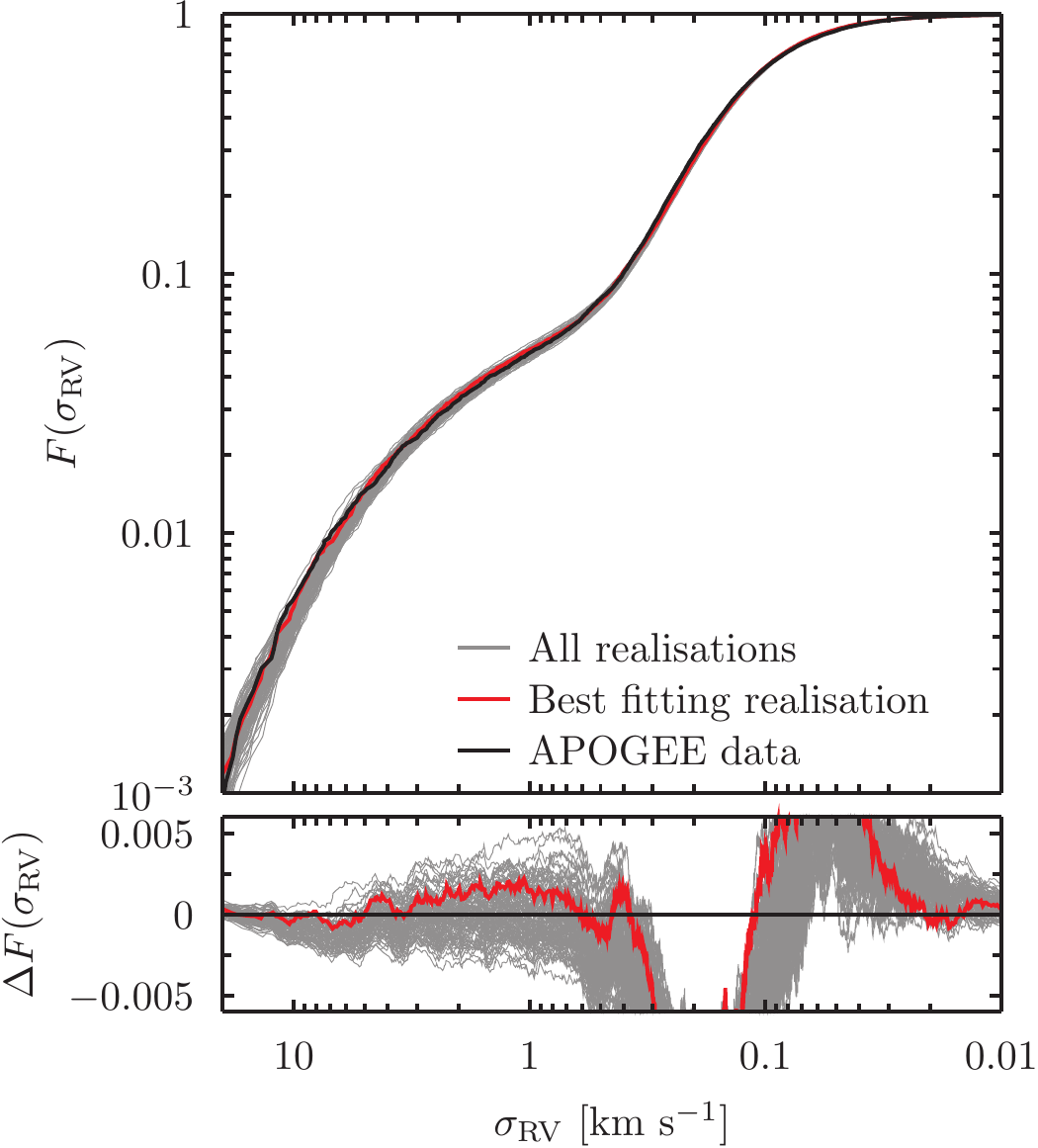}
\caption{
{\it Top panel:} Cumulative radial velocity scatter distributions $F(\sigma_{\rm
RV})$ for the best-fitting model.  The black line shows the APOGEE data in the bin
around solar metallicity.  The grey lines show 200 realisations of the model
with the best fitting parameters ($f_{\rm b,0}=0.35$, $\mu_s=-0.71$ and
$\sigma_s=0.29$).  The best fitting of the 200 realisations is shown in red.
{\it Bottom panel:} The same data plotted as as the difference between the model
and observational data.  The relatively poor fit at low RVs 
($\sigma_{\rm RV}\lesssim 0.5\,\kms$) is because the intrinsic scatter is not
exactly fit by a single log-normal distribution.
}
\label{fig:bestFitCumulative}
\end{figure}

Having done this we calculate a goodness-of-fit parameter $Q$ from the
cumulative distributions of scatter in the APOGEE data, $F_{\rm obs}(\sigma_{\rm
RV})$, and in the model $F_{\rm mod}(\sigma_{\rm RV})$, as

\begin{equation}
Q = \int_{0.01\,\kms}^{20\,\kms}\left(\log F_{\rm obs} - \log F_{\rm
mod}\right)^2 \ud \log \sigma_{\rm RV},
\end{equation}

where the cumulative fraction runs from high velocities to low velocities.  This
form for $Q$ is chosen to emphasise the high-scatter objects (i.e. the binaries)
at the expense of the low-scatter (i.e. single) stars.  The limits of the
integral are chosen to exclude a handful of outliers that make the measurement
noisy.  Using this definition for $Q$ we find a best-fit model by minimising $Q$
with respect to the model parameters $f_{\rm b,0}$, $\mu_s$ and $\sigma_s$.  For
small values of $Q$ (well-fitting models) the value is dominated by random
scatter; in these cases we calculate 200  realisations of our model with the
same parameters and choose the median value of $Q$.  The results are shown in
Figure~\ref{fig:bestFitCumulative}.  The model including both binary and single
stars reproduces the main features and shape of the distribution well.  This
suggests that the major cause of high-amplitude velocity scatter in the APOGEE
sample is indeed the presence of binary stars.

\begin{figure}
\includegraphics[width=\columnwidth]{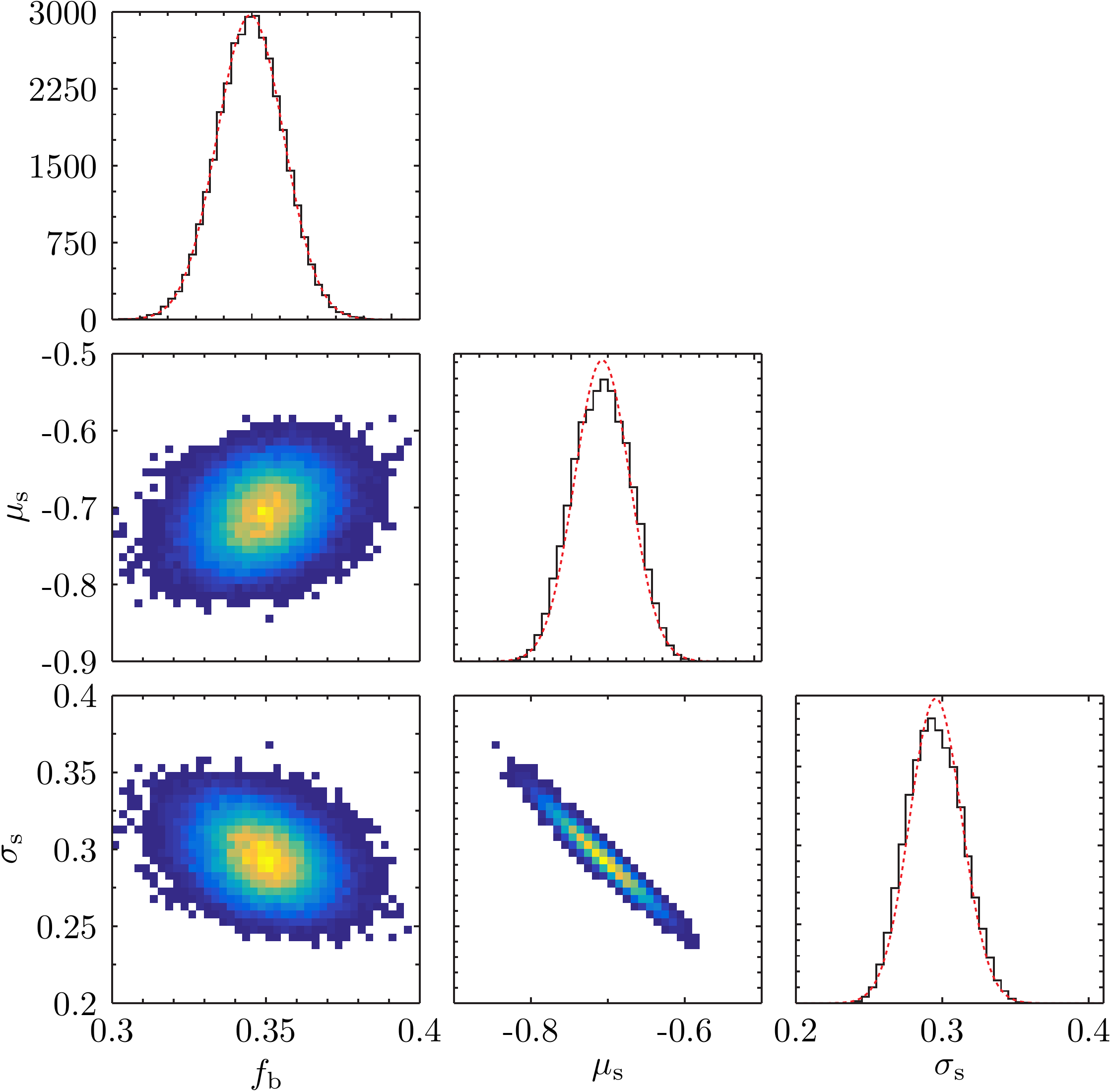}
\caption{Marginalised posterior probability density distributions for the
parameters of our model fit.  The parameters are the binary fraction at birth,
$f_{\rm b,0}$, and the mean and standard distribution of the RV scatter
intrinsic to the observations, $\mu_s$ and $\sigma_s$.  The red dashed lines
represent Gaussian fits to the one-dimensional distributions.  The calculations
were carried out for the metallicity bin centred on solar metallicity; see
Section~\ref{subsec:frequency} for details.
}
\label{fig:cornerPlot}
\end{figure}

Having derived the best fit values we apply the technique of approximate Bayesian
computation as presented by \citet{Turner12}.  We assume flat priors on all
three parameters and calculate $Q$ for a grid of models that span the best-fit
values.  Realisations of the model that have $Q$ less than a critical value
$Q_{\rm crit}$ then sample the posterior distributions of our parameters
directly.  We find that the process has converged at $Q_{\rm crit}\approx 0.02$.
Plots of the inferred one-dimensional posterior probability distributions of the
parameters and two-dimensional covariance distributions for the bin centred on
solar metallicity are shown in Figure~\ref{fig:cornerPlot}.  Once we have
marginalised over the nuisance parameters $\mu_s$ and $\sigma_s$ we find that
the posterior probability distribution of the initial binary fraction $f_{\rm
b,0}$ is well described by a Gaussian, and that for solar metallicity $f_{\rm
b,0} = 0.35\pm 0.01$.  

\subsection{Variation of binary fraction with metallicity}
\label{sec:Zvar}

We repeat the above analysis for the metallicity bins immediately above and
below Solar metallicity.  The results for all three bins are shown in
Table~\ref{tab:resultsMet}.   We reproduce the declining trend in binary
fraction with metallicity found by other authors in the APOGEE sample
\citep{Badenes17} and in other samples such as LAMOST or SEGUE
\citep{Yuan15,Gao17}. The binary fractions deduced from the LAMOST and SEGUE
samples are consistent with our results within the uncertainties.  Both samples
show the same trend: the fraction of binaries is smaller for stars of redder
colours and higher metallicities.  On the other hand \citet{Hettinger15} found
that, for F-type stars in the Milky Way, metal-rich stars were 30\% more likely
to have companions with periods shorter than 12 days than metal-poor stars.  Our
sample is dominated by red clump and RGB stars, which, by the time APOGEE
observes them, have evolved to be too large to fit in such a close binary.
Hence the binaries discussed by  \citet{Hettinger15} would not be present in our
sample or in our model output since they would have interacted before either
star reached the giant branch.

\begin{table}
 \caption{Effect of metallicity on the inferred initial binary fractions.}
 \begin{tabular}{rccc}
    \hline
  [Fe/H]     &  $f_{\rm b,0}$     &     $\mu_s$     &     $\sigma_s$     \\
  \hline          
  -0.30      &  $0.37\pm 0.01$  &     $-0.64\pm0.05$ &  $0.31\pm0.02$  \\
  0.00       &  $0.35\pm 0.01$  &     $-0.71\pm0.04$ &  $0.29\pm0.02$  \\
  0.17       &  $0.30\pm 0.01$  &     $-0.74\pm0.05$ &  $0.29\pm0.03$  \\
    \hline
 \end{tabular}
\label{tab:resultsMet}
\end{table}

\subsection{The effect of initial conditions}

\begin{table}
 \caption{Effect of binary initial conditions on inferred binary fraction at
 Solar metallicity}
 \begin{tabular}{lcccl}
    \hline
    Model set           &     $f_{\rm b,0}$     &     $\mu_s$        &     $\sigma_s$  &  Odds         \\
                        &                       &                    &                 &  ratio        \\
    \hline                                                                      
    {\tt standard}      &     $0.35\pm 0.01$    &  $-0.71\pm0.04$    &  $0.29\pm0.02$  &  1.0          \\
    {\tt m2fromIMF}   &                       & No good fit        &                 &  $<10^{-3}$   \\
    {\tt qGamma0.3}     &     $0.34\pm 0.01$    &  $-0.72\pm0.04$    &  $0.30\pm0.02$  &  1.8          \\
    {\tt qGamma1.0}     &     $0.32\pm0.01$     &  $-0.73\pm0.03$    &  $0.31\pm0.01$  &  3.5          \\
    {\tt periodR10}     &     $0.37\pm 0.01$    &  $-0.70\pm0.04$    &  $0.29\pm0.02$  &  0.38         \\
    {\tt flatEcc}       &     $0.33\pm 0.01$    &  $-0.70\pm0.04$    &  $0.29\pm0.02$  &  0.33         \\
    {\tt old}           &     $0.32\pm0.01$     &  $-0.72\pm0.03$    &  $0.30\pm0.01$  &  2.0          \\
    {\tt young}         &     $0.12\pm0.01$     &  $-0.71\pm0.03$    &  $0.30\pm0.01$  &  1.2          \\
    \hline
 \end{tabular}
\label{tab:binFracInitCond}
\end{table}

We repeat the analysis from Section~\ref{subsec:frequency} for the different set
of binary properties given in Table~\ref{tab:models}; see
Table~\ref{tab:binFracInitCond} for the results.  Good fits are obtained for all
models other than the {\tt m2FromIMF} model, in which we draw the masses of both
stars independently from the IMF. The values of $\mu_s$ and
$\sigma_s$ are independent of the binary model used within the uncertainties.
This is expected, since they are properties that primarily affect the single
stars where they are the only source of observable velocity scatter, and hence 
their constancy is a necessary condition for the models to be good fits.  The
values of $f_{\rm b,0}$ do, however, change depending on the model used.  This
is because different sets of assumptions about the binary properties lead to
different fractions of binaries evolving into the range of parameter space where
they contribute to the scatter in APOGEE observations.  The final column shows
the odds ratio, calculated as the fraction of the three-dimensional phase space
contributing systems to the posterior distribution at a given cut in $Q$,
relative to the {\tt standard} model set.  Were the different sets of initial
conditions equally likely {\it a priori} this could be interpreted as their
relative likelihood, and hence gives an indication of how strongly the different
possibilities are supported by the data.

\begin{figure}
\includegraphics[width=.9\columnwidth]{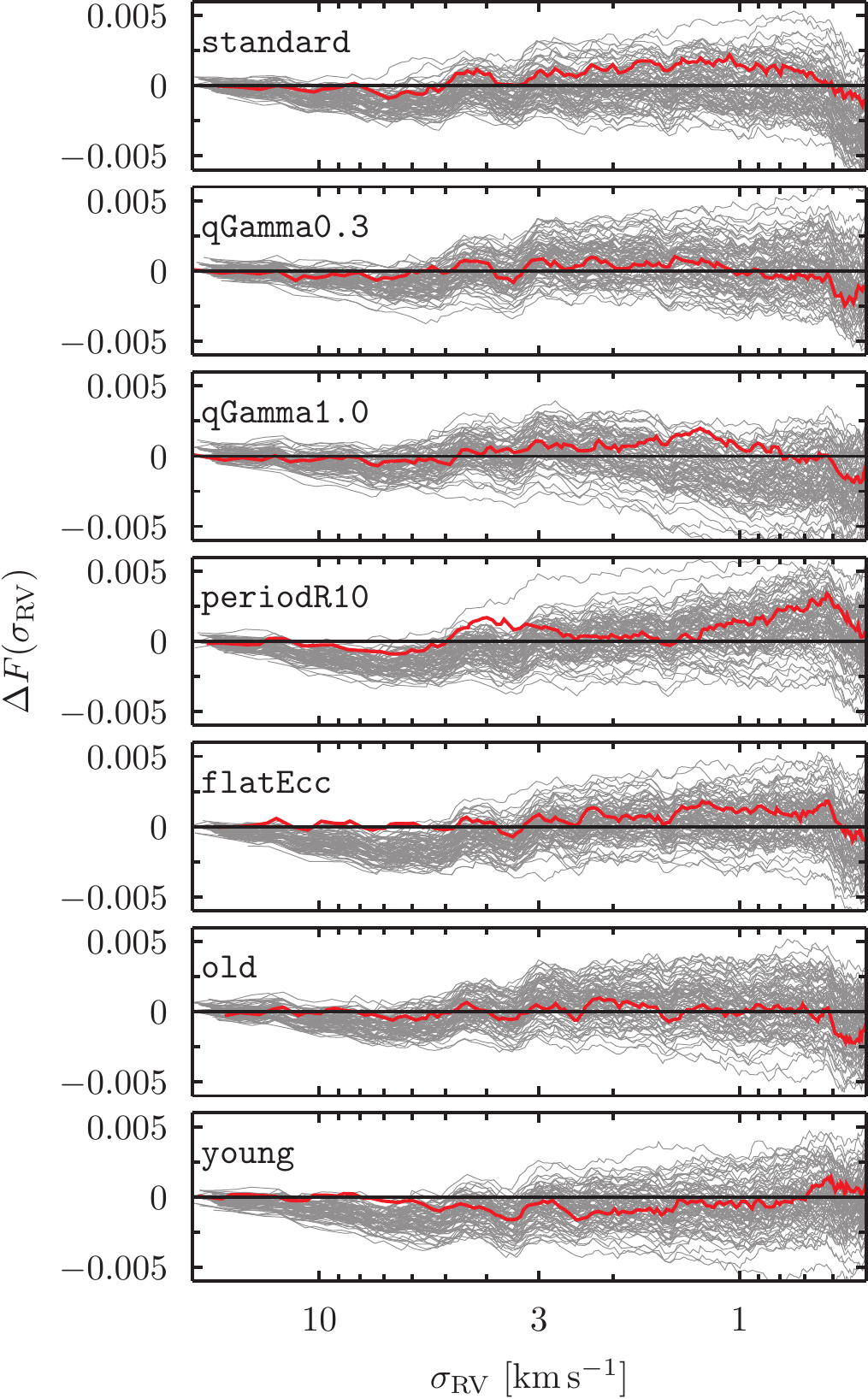}
\caption{
Difference between the cumulative distribution of the APOGEE observations and
that of the model, $\Delta F(\sigma_{\rm RV}) = F_{\rm mod}(\sigma_{\rm RV}) -
F_{\rm obs}(\sigma_{\rm RV})$, as a function of the RV scatter
$\sigma_{\rm RV}$.  Each panel shows a different set of initial conditions as
given in Table~\ref{tab:models}.  The grey lines show 100 realisations of the
model, each with the best-fit parameters taken from
Table~\ref{tab:binFracInitCond}, and the red line shows the best fitting of
these models for each parameter set.  The black line shows the offset for the
data, which is zero by definition.  Only the part of the plot up to $0.5\,\kms$
is shown since the lower velocities are dominated by single stars and binaries
with negligible reflex motion.  The distribution for the {\tt m2fromIMF} initial
conditions is omitted since no good fit was found.
}
\label{fig:cumulativeDiffs}
\end{figure}

In Figure~\ref{fig:cumulativeDiffs} we plot the differences between the observed
and model cumulative scatter distributions, $\Delta F(\sigma_{\rm RV})$, for
the different models.  For each set of binary parameters we show 100
realisations of the best fit model given in Table~\ref{tab:binFracInitCond},
with the best fitting of those 100 highlighted in red.  Each of the models shows
rather similar behaviour, and changing the input parameters within these
reasonable bounds of disagreement shows no significant improvement to the fit.
The only exception is the model where we draw both stars independently from the
IMF, which is very strongly disfavoured.  This implies that the binaries that we
are able to constrain using this model come from a limited sub-set of the
initial binary populations, a hypothesis that we now go on to test.

\subsection{Which binaries does the APOGEE data constrain?}

\begin{figure*}
   \includegraphics[width=.8\textwidth]{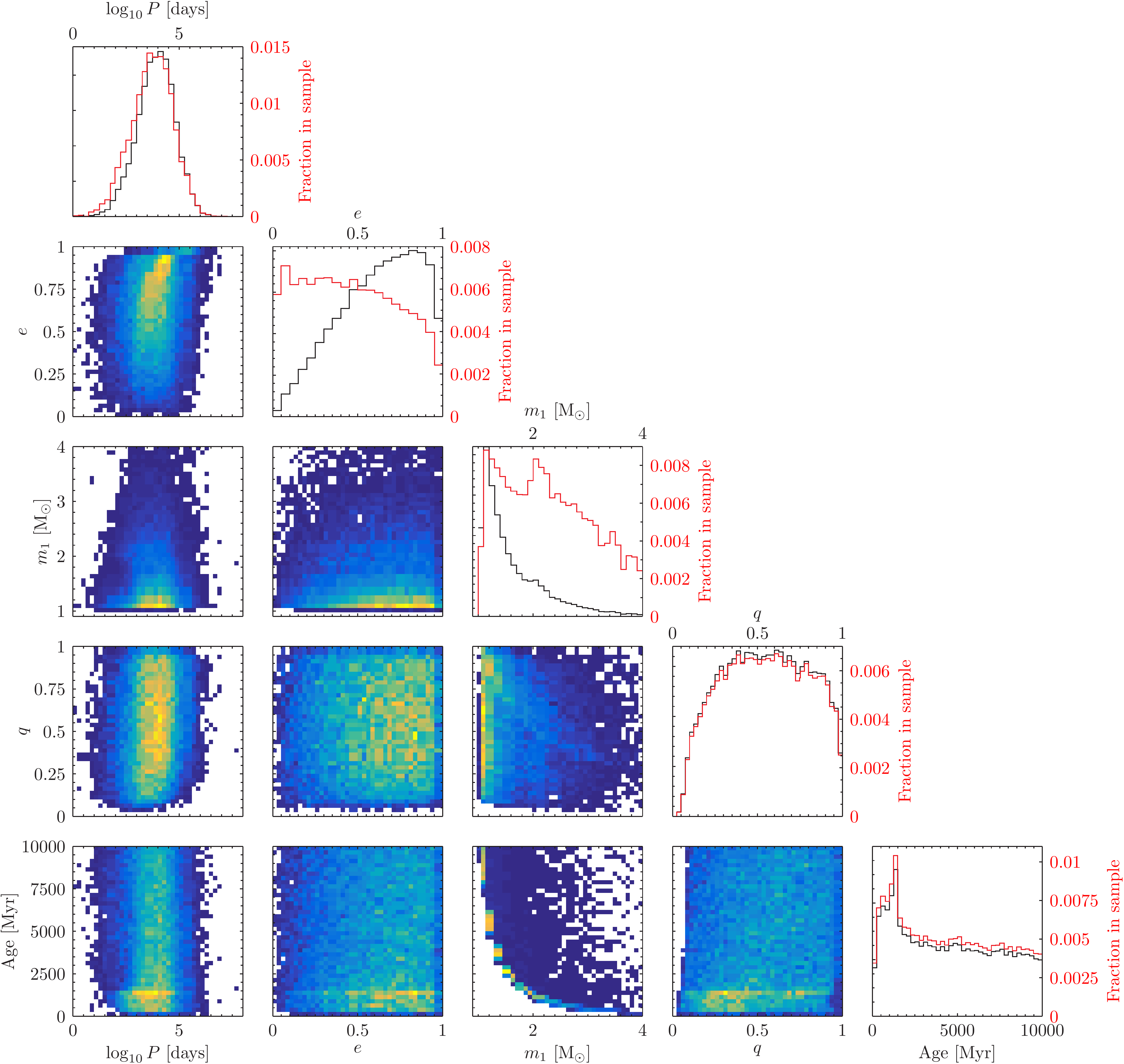}    
\caption{Distributions of birth properties of the binaries in our model which we
predict would be found in APOGEE with a projected RV scatter
$\sigma_{\rm RV}>0.5\,\kms$.  The quantities plotted, left to right along the
abscissa, are:
the log of the binary period in days; the eccentricity; the mass of the initially
most massive star, $m_1$, in $\msun$; the mass ratio $q$; and the age in Myr.
The colour maps show two-parameter correlations.  The top panel in each column
shows the distribution of that property in the sample.  The black line shows the
relative number of stars at different property values that meet the selection
criteria and have  $\sigma_{\rm RV}>0.5\,\kms$.  The red line (with labels on
the right-hand axis) shows the fraction of the complete initial sample of
binaries that meet our selection criteria to be observed by APOGEE and have
$\sigma_{\rm RV}>0.5\,\kms$.  Where this has a different shape to the black line
it is because the distribution of stars in the initial sample is significantly
non-uniform across the relevant parameter range (e.g. for the eccentricity).
}
\label{fig:whichBinaries}
\end{figure*}

Figure~\ref{fig:whichBinaries} shows the birth properties of binaries in our
sample of synthetic APOGEE observations that have $\sigma_{\rm RV}>0.5\,\kms$.
This value was chosen as the approximate point where the cumulative distribution
goes from being dominated by non-binary RV scatter to RV scatter produced by the
binary orbit.  Hence these are the binaries that our analysis is sensitive to.
Binaries with other initial properties also appear in our sample, but have
insufficient velocity scatter to be clearly separated from single stars.
Roughly half of the binaries in our synthetic observations lie above this limit;
binaries below this limit cannot be detected through RV scatter
alone.

Figure~\ref{fig:whichBinaries} shows that the main quantity that determines
whether a binary is observed to have $\sigma_{\rm RV}>0.5\,\kms$ is the initial
period. The binaries that appear mostly have periods between $10^3$ and $10^4$
days.  They are thus sufficiently wide that neither star has filled its Roche
lobe before becoming a giant, but close enough that they still have a
sufficiently high RV that $\sigma_{\rm RV}>0.5\,\kms$.  Appearance
is largely independent of initial eccentricity, except at very high eccentricity
where the separation at pericentre is greatly reduced.  The visible correlation
between the mass of the initially most massive star, $m_1$, and age is because,
for the most massive star to appear in the APOGEE sample, it must have evolved
to the giant branch but no further.  The other effect that is visible is a weak
preference for young ages.  This may be because the stars that are evolving on
to the giant branch at these ages never develop degenerate cores, and hence
never grow as large on the first giant branch.  Hence a larger range of orbits
can survive to the horizontal branch, where the star is longer-lived in the
APOGEE-observable sample.

These effects explain the differences in binary fractions seen in
Table~\ref{tab:binFracInitCond}.  At young ages there are a higher fraction of
binaries visible in the model, so the young sample requires a smaller binary
fraction to reproduce the observed scatter.  The \citet{Raghavan10} distribution
has a smaller fraction of binaries in the period range where they influence our
model so a slightly larger binary fraction is required.  The other binary
properties have little effect on the population of binaries with $\sigma_{\rm
RV}>0.5\,\kms$ and hence do not change the binary fraction that we infer.


\section{Conclusions and discussion}
\label{sec:Conclusions}


We have synthesised realistic stellar populations, composed of a mix of single
and binary stars, using the rapid binary-star evolution code {\sc bse}.  As
an application we model the effect of binary stars on observations of Galactic
disc stars in the Apache Point Observatory Galactic Evolution Experiment (APOGEE
DR\,12).  We compare the radial velocity (RV) scatter in representative samples at
different metallicities and find that the scatter found at velocities above
$0.5\,\kms$ in the APOGEE data is adequately reproduced by the effects of
stellar binaries.  We find correlations between the stellar atmospheric
parameters and RV scatter that are similar to those in the APOGEE
sample.  The APOGEE sample is sensitive to the full range of binary
eccentricities and mass ratios, primary star masses between the turnoff mass and
a few solar masses, and initial binary periods between 3 and 3000\,years.

Using this approach we constrain the birth binary fraction of solar metallicity
binaries to be $f_{\rm b,0} = 0.35\pm 0.01$.  This is consistent with the study
of solar-neighbourhood, solar-type binaries by \citet{Raghavan10}, who find
$f_{\rm b,0}=0.33\pm 0.02$.  Our results are relatively insensitive to
reasonable choices of initial binary properties.  Both the period distributions
of \citet{Raghavan10} and \citet{Duquennoy91} are consistent with the data, with
a slight preference for the latter.  Similarly flat and thermal eccentricity
distributions both lead to good fits, with a slight preference for a thermal
distribution.  The mass ratio distribution is somewhat better fit with a mild
bias towards equal mass ratios; a power-law with index $\gamma=1$ fits better
than a flat distribution, and choosing both stars independently from the IMF is
ruled out.  We are largely insensitive to the stellar age distribution.  The
majority of stars observed by APOGEE are old, and hence significant changes in
the age distribution only translate to small changes in the distribution of
initial stellar masses in the observed sample.  Hence assuming a flat age
distribution with a broad range of ages is sufficient.

An extension of this work would be to fit the properties of the initial binary
population simultaneously with the binary fraction.  Whilst such an analysis
would be possible, we have not attempted it here for two reasons.  First, it
would require either synthesis of a new sample for each model value explored, or
synthesis of a much larger population of binaries which could then be
re-sampled.  Either option would be much more computationally intensive.
Secondly, many of the properties of the population -- e.g. the width of the
distribution of orbital periods -- are poorly constrained by our data and hence
little would be gained by fitting for them.  Instead of fitting for these
parameters we have chosen to explore a range of reasonable values.  The downside
of this approach is that the uncertainties that we present in
Table~\ref{tab:binFracInitCond} are underestimates.  A better estimate of the
uncertainty on our measurement of the initial binary fraction can be seen in the
scatter between different reasonable choices of initial conditions, which is of
the order of a few percentage points.

We also study the effect of metallicity, and reproduce the trend of a declining
binary fraction with increasing metallicity.  This is in qualitative agreement
with the hydrodynamic star-formation models of \citet{Machida09}, whose results
indicate a higher binary frequency in lower metallicity gas. On the contrary,
the recent results obtained from radiation hydrodynamical simulations by
\citet{Bate14} suggest that stellar binaries are formed mainly by gravitational
collapse, which is highly insensitive to the dust content of the protostellar
disk.  More comprehensive studies, particularly of large samples of dwarf stars
where we are more sensitive to shorter-period binaries, are necessary to resolve
this question.

We do not consider the lowest metallicity stars present in the
APOGEE sample (i.e. those with [Fe/H]$< -0.5$).  The reasons for this are two-fold.
First, relatively few stars meet our criteria for inclusion, so the constraints
available on the binary population are weaker.  Secondly, we start to see an
increasing population of sub-luminous, cold stars with high velocity dispersions
in the APOGEE data, similar in appearance to Algols.  Very few such stars are
present at solar metallicity, but at lower metallicities they become quite
common, a trend that is not well reproduced by our {\sc bse} models.  We leave
investigation of these binaries to a subsequent study.

It is unsurprising that the binary reflex motion dominates the RV
scatter at higher velocities.  Other astrophysical sources of RV
scatter include higher-order stellar multiples, planetary systems, stellar
surface oscillations and varying star-spot cover.  There is also scatter induced
from errors in spectral analysis and instrumental effects.  All of these
contributions are expected to be smaller than the binary contribution, and this
fits with the detailed predictions of our models.  Therefore, although we are not
sensitive to all binaries, we find that stellar surveys with multiple
observations per star are sensitive to the overall binary population.  Whilst we
have focussed on giants in this paper, samples involving dwarfs will be
sensitive to closer binaries, which do not in general form binaries as the stars
interact before the reach the giant branch.  We intend to investigate the local
APOGEE dwarf sample in a follow-up paper.

Our analysis is easily adapted to other surveys with different observation
schedules, selection functions, wavebands, etc.  Careful consideration of the
binary population is necessary to quantify the effects of unresolved binary
systems on measurements derived from stellar spectroscopy.  An analysis such as
this one is useful to allow the effects of binaries on the survey science to be
quantified and the survey design optimised to reduce the effects of binaries.
We plan to expand a similar analysis to other existing spectroscopic surveys,
large and small, and also surveys currently in planning and development such as
WEAVE and 4MOST.

{\small
\section*{Acknowledgements}
The authors thank Will Farr, Jarrod Hurley and Christopher Tout for helpful
discussions.  E.S.~acknowledges support from the European Social Fund via the
Lithuanian Science Council grant No. 09.3.3-LMT-K-712-01-0103.  The authors
thank the Knut and Alice Wallenberg Foundation for support through the project
grants ``The New Milky Way'' and ``IMPACT''.  R.P.C.~was supported by funds from
the eSSENCE Strategic Research Environment. Simulations discussed in this work
were performed on resources provided by the Swedish National Infrastructure for
Computing (SNIC) at the Lunarc cluster, funded in part by the Royal Fysiographic
Society of Lund.  Funding for SDSS-III has been provided by the Alfred P. Sloan
Foundation, the Participating Institutions, the National Science Foundation, and
the U.S. Department of Energy Office of Science. The SDSS-III web site is {\tt
http://www.sdss3.org/}.

}

\bibliographystyle{mnras}
\bibliography{bse}
\bsp	

\label{lastpage}
\appendix

\section{Additional Data}
\label{app:BSEparams}

The binary evolution algorithm described by \citet{Hurley02} has a large number of tunable parameters. 
In Table\,\ref{tab:data} we list the values of the input parameters not described in more detail in Section~\ref{sec:methods}.
 \begin{table*}
	\centering
    \caption{Input parameters for binary evolution in the {\sc bse} algorithm.}
    \begin{tabular}{lll}
    \hline
    \noalign{\smallskip}
Parameter & Value &  Meaning \\
    \noalign{\smallskip}
    \hline
    \noalign{\smallskip}

$\eta	$	& 0.5  	& 		Reimers mass-loss coefficient\\
$B_{W}$	& 0.0  	&		Binary enhanced mass loss parameter\\
$He_{W}$	& 1.0  	&		Helium star mass loss factor\\
$\alpha_{CE}$	& 3.0	&		Common-envelope efficiency parameter\\
$\lambda$	& 0.5	  	&  		Binding energy factor for common envelope evolution\\
$\sigma_{k}$	& 190 &		Dispersion in the Maxwellian for the SN kick speed (in km/s)\\
$\beta_{W}$& 0.125	&		Wind velocity factor\\
$\upmu_{W}$& 1.0	&		Wind accretion efficiency factor\\
$\alpha_{W}$& 1.5	&		Bondi-Hoyle wind accretion factor\\
$\upepsilon$	& 0.001&		Fraction of accreted matter retained in nova eruption\\
Eddfac	& 1.0		&		Eddington limit factor for mass transfer\\
$\Gamma$& -1.0	&		Angular momentum factor for mass lost during Roche lobe overflow\\
    \noalign{\smallskip}
    \hline
     \end{tabular}
       \begin{tabular}{l}
   {\it{Note.}} The parameter values are the defaults, as described in \citet{Hurley02}.
    \end{tabular}
\label{tab:data}
\end{table*}
\end{document}